\documentclass[12pt,preprint,natbib209]{aastex}

\shorttitle{Fermi Bubbles from star-forming Regions }
\shortauthors{de Boer, Weber}
\begin{document}


\title{Fermi Bubbles and Bubble-like emission from the Galactic Plane}
\author{Wim de Boer and Markus Weber}
\affil{Institut f\"ur Experimentelle Kernphysik, Karlsruhe Institute of Technology, P.O. Box 6980, 76049 Karlsruhe, Germany}
\email{wim.de.boer@kit.edu; markus.weber2@kit.edu }


\begin{abstract}
The diffuse gamma-ray sky revealed 'Bubbles' of emission above and below the Galactic Plane symmetric around the centre of the Milky Way with a height of 10 kpc in both directions.  At  present  there is no convincing explanation for the origin. To understand the role of the Galactic Centre (GC) one has to study the Bubble spectrum inside the disc, a region which has been excluded from previous analysis because of the large foreground.  From a novel template fit, which allows a simultaneous determination of  the signal and foreground in any direction, we find that  bubble-like emission  is not only found in the halo, but in the Galactic Plane as well with a width in latitude coinciding with the molecular clouds. The  longitude distribution has a width corresponding to  the Galactic Bar with an additional contribution from the Scutum-Centaurus arm. The energy spectrum of the Bubbles  coincides with the predicted contribution from  CRs trapped in sources (SCRs). Also the energetics fits  well. Hence, we  conclude that the bubble-like emission has a hadronic origin, which arises from SCRs and the Bubbles in the halo arise from hadronic interactions in advected gas. Evidence for advection  is provided by the ROSAT X-rays from hot gas in the Bubble region. 
\end{abstract}

\section{Introduction}

The FERMI-LAT gamma-ray telescope \citep{Atwood:2009ez} has surveyed the  gamma-ray sky  at energies between 100 MeV and 100   GeV or even above with unprecedented precision.
The main contributions  to the gamma-rays are well understood: $\pi^0$ production by nuclear interactions combined with a smaller leptonic component: bremsstrahlung (BR) and inverse Compton (IC) scattering of leptons on the photons of the interstellar radiation field (ISRF) \citep{Strong:2007nh}.   The Fermi  Bubbles were discovered as an excess over the expected background \citep{Su:2010qj}.  The  origin of the Bubbles  is unclear and many proposals have been made, ranging from hadronic cosmic rays (CRs) interacting with hot gas in the halo \citep{Crocker:2010dg}, to star bursts \citep{Biermann:2009td}, to AGN activity in our Galaxy \citep{Guo:2011eg,Guo:2011ip,Yang:2012fy,Cheng:2011xd},  to originating from dark matter annihilation   \citep{Dobler:2011mk}.

\begin{figure}
\centering
\includegraphics[width=0.45\textwidth]{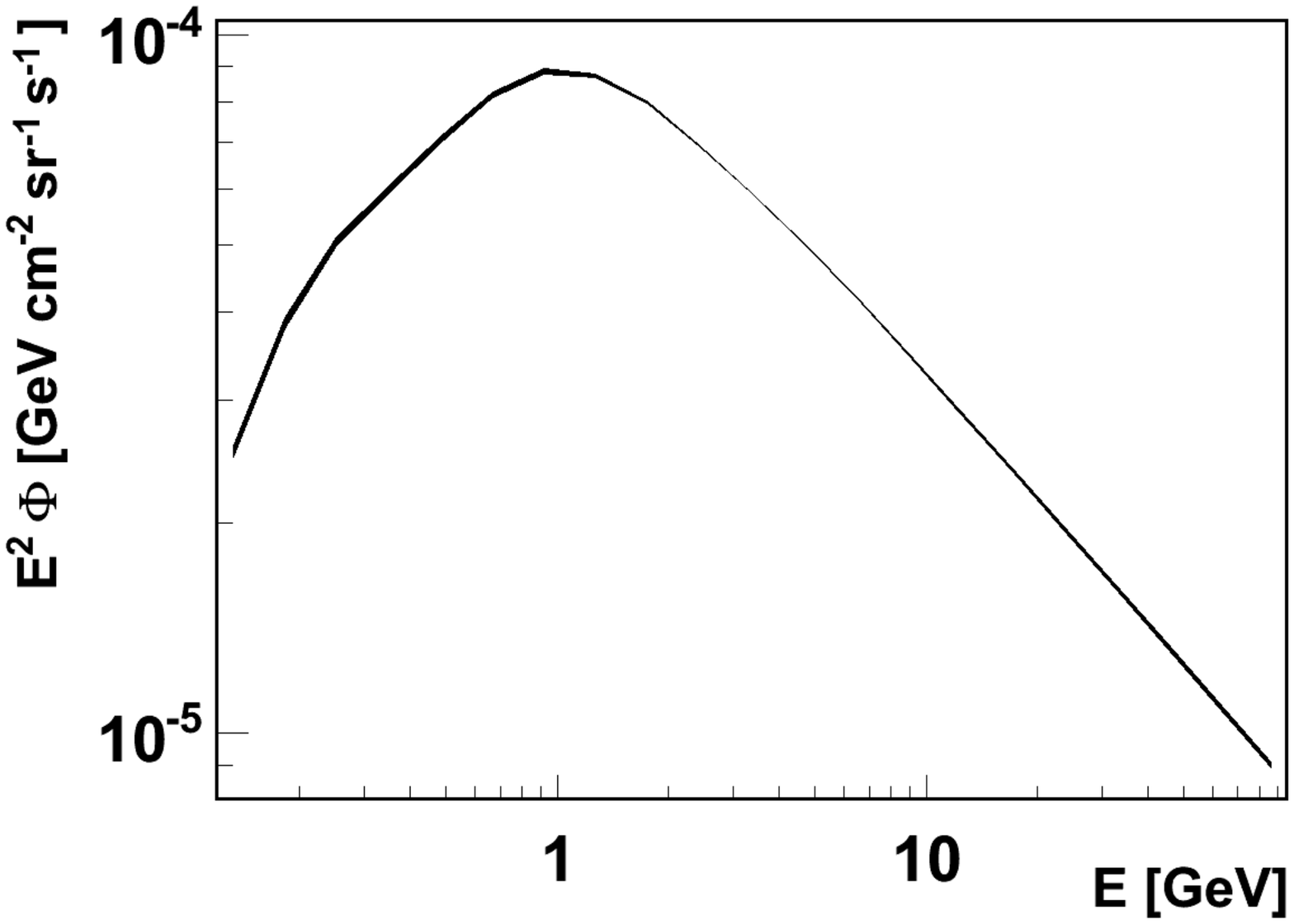}
\includegraphics[width=0.45\textwidth]{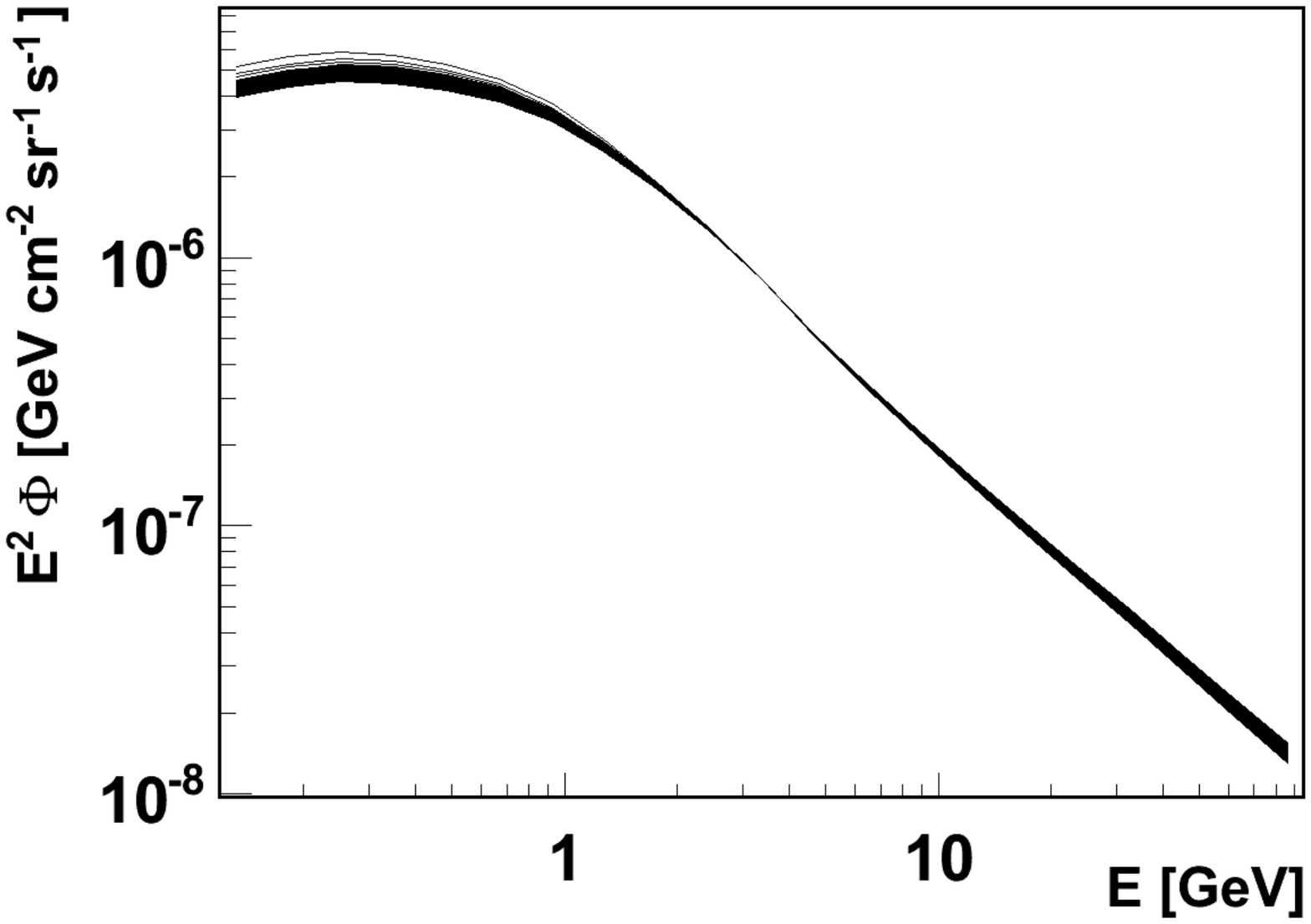}\\[-30mm]
\hspace*{0.05\textwidth}(a)\hspace*{0.5\textwidth} (b)\\[-20mm]
\includegraphics[width=0.45\textwidth]{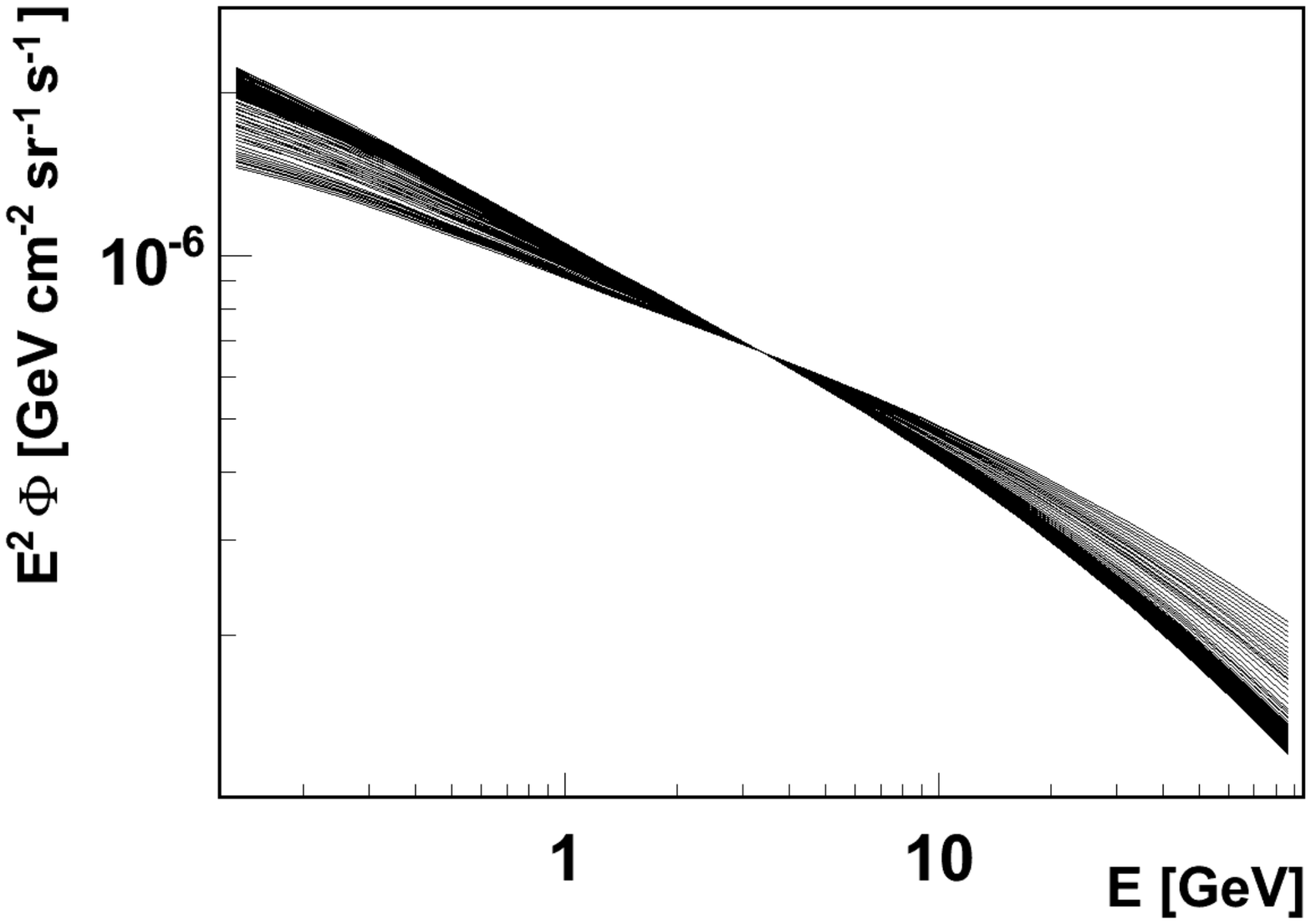}
\includegraphics[width=0.465\textwidth]{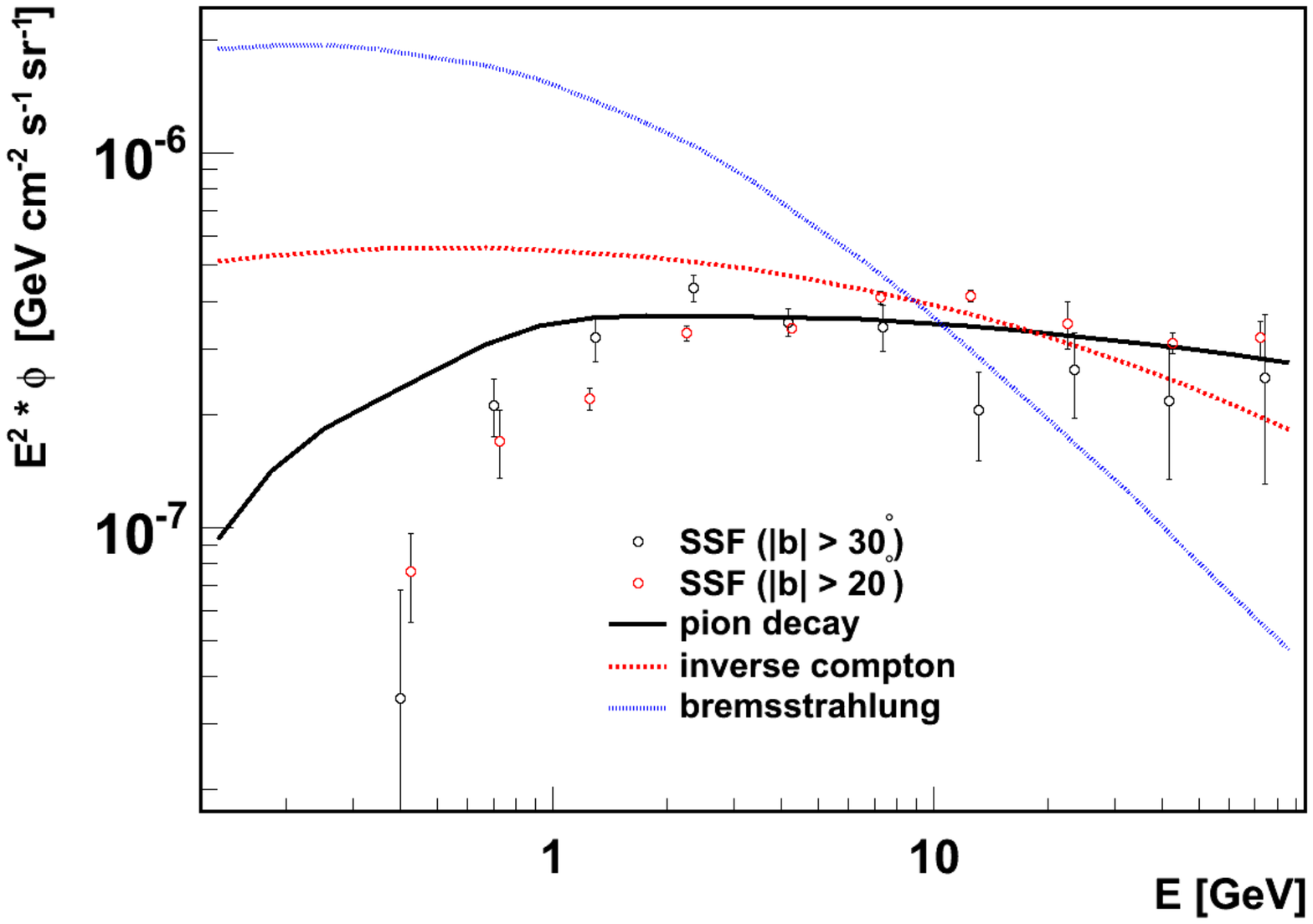}\\[-30mm]
\hspace*{0.05\textwidth}(c)\hspace*{0.5\textwidth} (d)\\[1mm]
\caption[]{Spectral templates for decays of $\pi^0$ mesons (a), bremsstrahlung (b) and inverse Compton scattering (c). Each panel has normalised  templates  superimposed for 1025 different directions. In panel (d) our Fermi Bubble template is shown as a black line, obtained from   $\pi^0$ production by freshly-accelerated protons. For comparison,   IC and BR 
templates for freshly-accelerated electrons are shown as well. The data points for $|b|>30^\circ$ and $|b|>20^\circ$ are from  \citet{Su:2010qj}  and \citet{Su:2013}, respectively.
 \label{f1}
}
\end{figure}
 
At present the morphology of the Bubbles has not been examined in the Galactic Plane, because of the large foreground, although such a determination could be helpful for the interpretation. Here we report a template fit to the Fermi data, which allows a simultaneous determination of the foreground and Fermi  Bubble signature in all sky directions, including the Galactic Plane. The idea is simple: one knows the energy spectra of the main contributions to the gamma-ray sky from accelerator experiments. If these foregrounds describe the data, one should be able to describe the gamma-ray spectra by a linear combination from the different foreground spectra (templates). The normalisation of each template is left free in a $\chi^2$ fit. Additional contributions will be apparent from a poor $\chi^2$. Adding the template needed for the Bubbles improved the $\chi^2$ in specific regions. With this method we reproduce the observed Fermi Bubbles in the halo, but find additionally bubble-like emission in star-forming regions in the Plane.
The Bubble template best describing the data corresponds to the hard proton CR spectra expected for shock wave acceleration by supernovae remnants \citep{Hillas:2005cs}, which already indicates that the Fermi Bubbles might be connected to CRs at the time when they are still trapped in sources, called "Source CRs" (SCRs) by \citet{2004ApJ...611...12B}. This observation is in line with the fact that propagation models, which do not include these SCRs, underpredict the gamma-ray production in the Plane \citep{FermiLAT:2012aa}. The observed excess has been interpreted by \citet{2013ApJ...777..149V} as evidence for SCRs, although the excess is comparable to the uncertainty from the propagation models. In this Letter we strengthen this interpretation by the novel template fitting method, which does not depend on the propagation model and allows for a  fine spatial resolution.

\section{Analysis}

We have analysed the diffuse gamma-rays in the energy range between 0.1 and 100 GeV using the diffuse class of the public  P7REP\_SOURCE\_V15 data collected from August, 2008 till July 2014 (72 months) by the Fermi Space Telescope \citep{Atwood:2009ez}. The data were analysed  with the recommended selections for the diffuse class using the  Fermi Science Tools (FST)  software \citep{FST}.
The sky maps were binned in longitude and latitude in 0.5$^\circ$x0.5$^\circ$ bins, which could later be combined at will. The point sources  from the second Fermi point source catalogue \citep{Fermi-LAT:2011iqa} have been subtracted using the gtsrc routine in the FST. The recommended  selection of events allows one to take the events from misidentified hadrons into account, which are part of the isotropic component provided by the Fermi software. 

Although the Fermi-LAT satellite has full sky coverage we concentrate on the region towards the GC where the Bubbles have been observed. We divided the sky in 41 3$^\circ$ bins in latitude and 25 5$^\circ$ bins in longitude, which leads to a coverage of 123 (125)$^\circ$ in latitude (longitude). This region-of-interest with 41x25 =1025 subcones was centred on the GC. 

The diffuse gamma-rays  trace the distribution of cosmic rays, since the CRs  collide sometimes with the gas or ISRF, which   leads to energetic gamma-rays \citep{Strong:2007nh}.  
The flux is proportional to the product of the CR densities, the "target densities" (gas or ISRF) and the cross sections, but  a template fit  lumps the product of these three factors into a single normalisation factor  for each gamma-ray component $k$, thus eliminating their individual uncertainties.

 The total flux in a given direction can be described as a linear combination of the various contributions with known energy spectra (templates):
 \begin{equation} |\Phi_{tot}>=n_1|\Phi_{\pi^0}>\  +\  n_2|\Phi_{BR}>\   +\  n_3|\Phi_{IC}>\  +\  n_4|\Phi_{Bubble}>\  +\  n_5|\Phi_{isotropic}>, \label{e2}\end{equation} where the normalisation factors $n_i$ determine the fraction of the total flux for a given contribution. The normalisations $n_i$  can be found from a $\chi^2$ fit, which tries to adjust the templates to best describe the data.  Since the number of data points in each spectrum for a given subcone (21 energy bins)  is large compared with  the number of free parameters ($n_i<5$) the fit is strongly constrained, thus allowing a precise determination of the various template contributions in each direction.

As test statistic we use the  $\chi^2$ function defined as:
 \begin{equation} \chi^2=\sum_{i=1}^{N=1025} \sum_{j=1}^{21}\left[\frac{\langle data(i,j)- \sum_{k=1}^{m\le5} n(i,k) \times template(i,j,k)\rangle^2}{\sigma(i,j)^2}\right], \label{e1}\end{equation}
where data(i,j) is the Fermi flux in direction $i$ for energy bin $j$, template(i,j,k) with normalisation $n(i,k)$ is the contribution of template $k$ out of a total of $m$ templates to data(i,j) and  $\sigma(i,j)$ is the total error of data(i,j), obtained by adding the statistical and systematic errors  in quadrature. The recommended systematic errors  in the Fermi Software on the total flux  are 10\% for gamma-ray energies below 100 MeV, 5\% at 562 MeV, and 20\% above 10 GeV. We used  a linear interpolation for energies in between.

\begin{figure}
\centering\vspace*{-20mm}
\includegraphics[width=0.4\textwidth,clip]{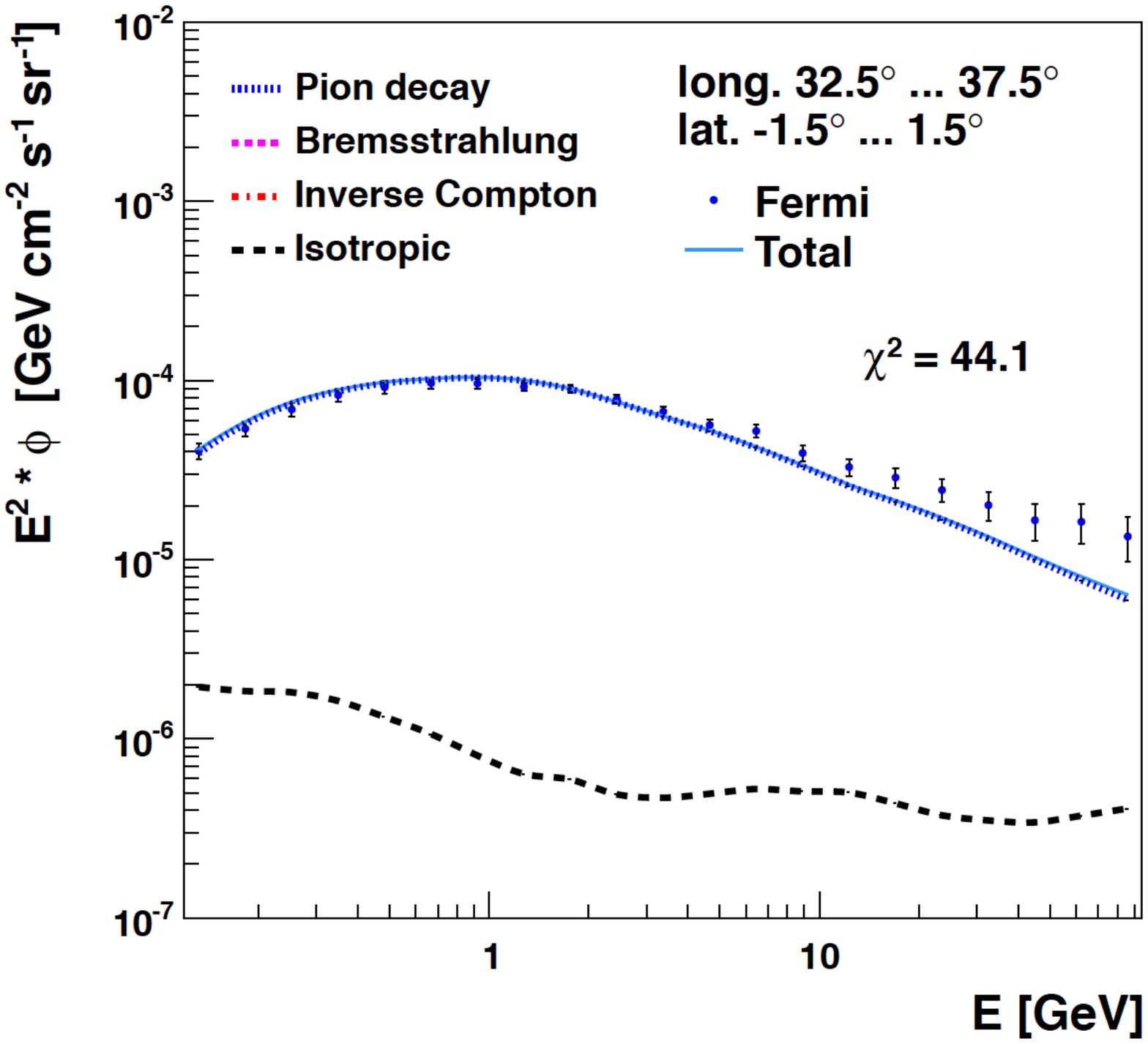}
\includegraphics[width=0.4\textwidth,clip]{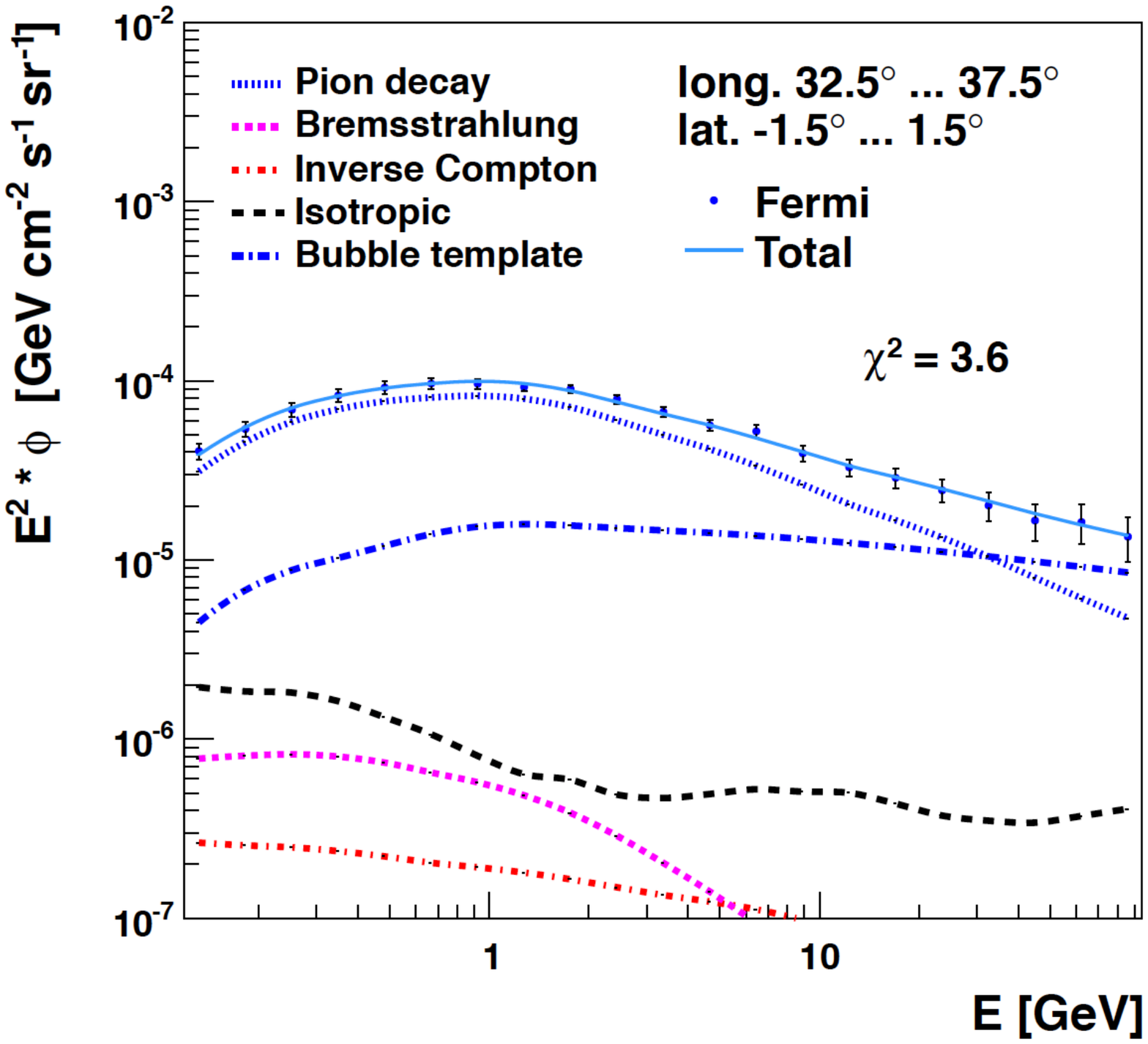}\\[-20mm]
\hspace*{0.03\textwidth}(a)\hspace*{0.35\textwidth} (b)\\[-18mm]
\includegraphics[width=0.4\textwidth,clip]{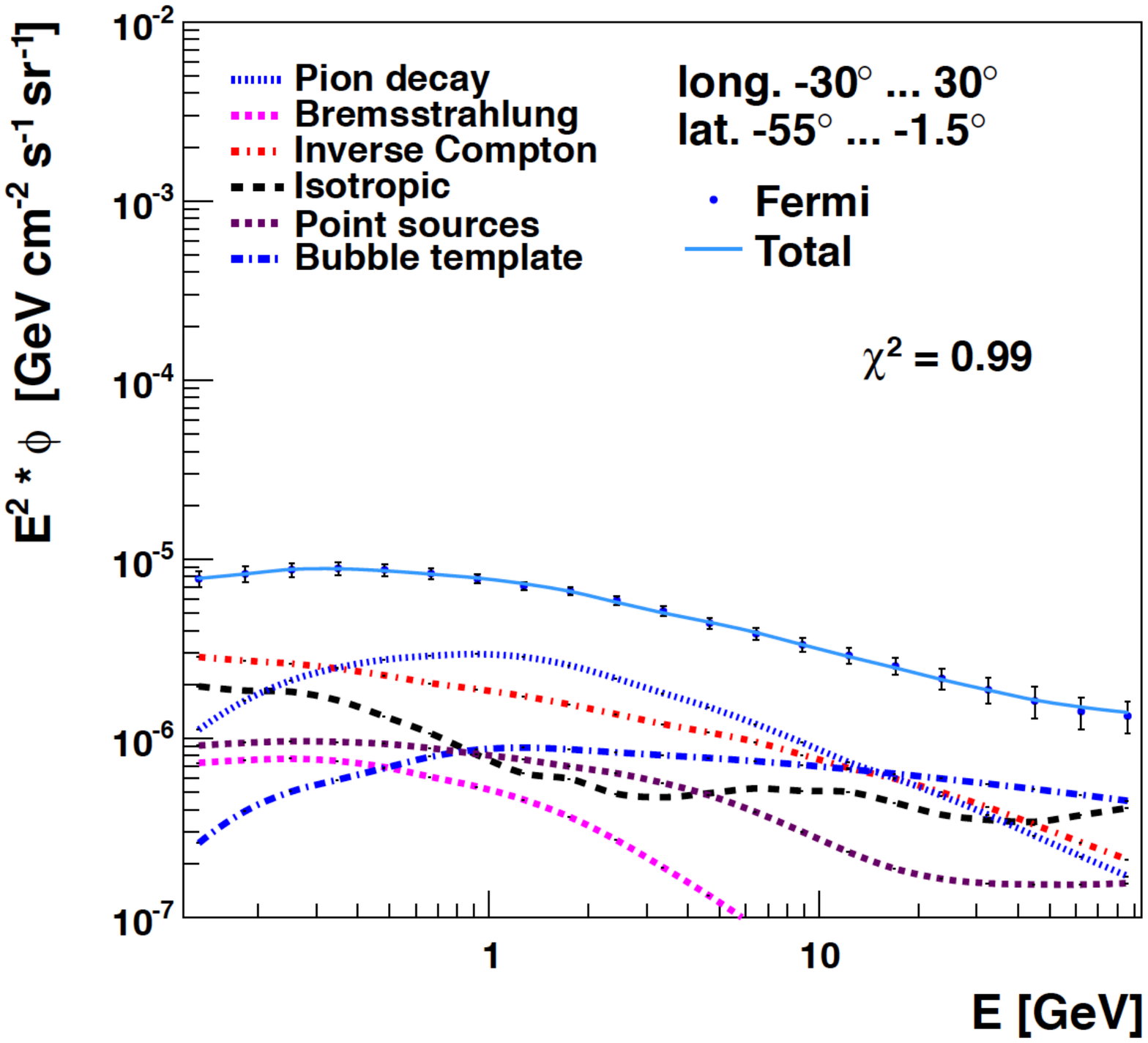}
\includegraphics[width=0.4\textwidth,clip]{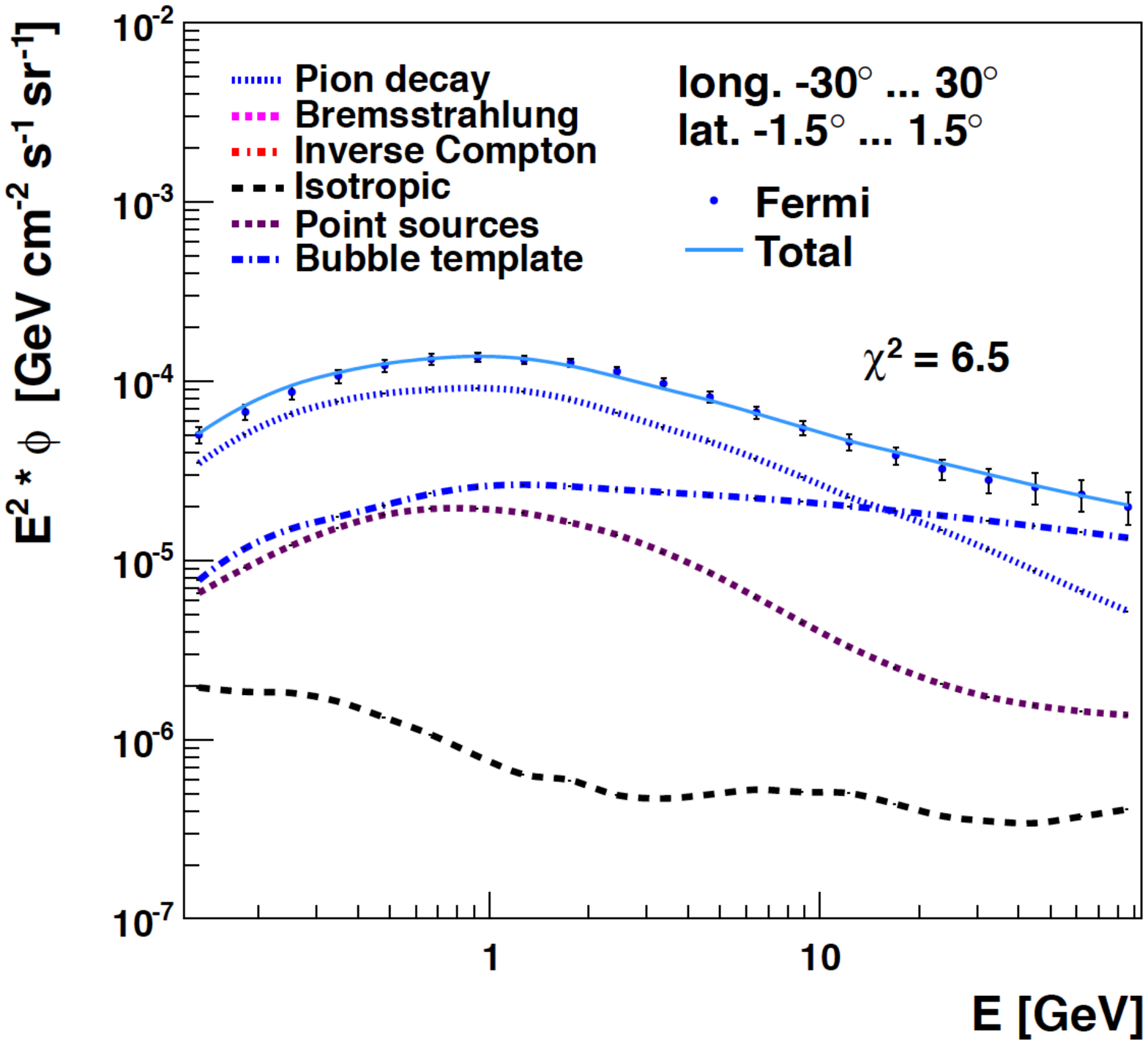}\\[-20mm]
\hspace*{0.03\textwidth}(c)\hspace*{0.35\textwidth} (d)\\[-12mm]\hspace*{2mm}
\includegraphics[width=0.385\textwidth,height=0.5\textwidth,clip]{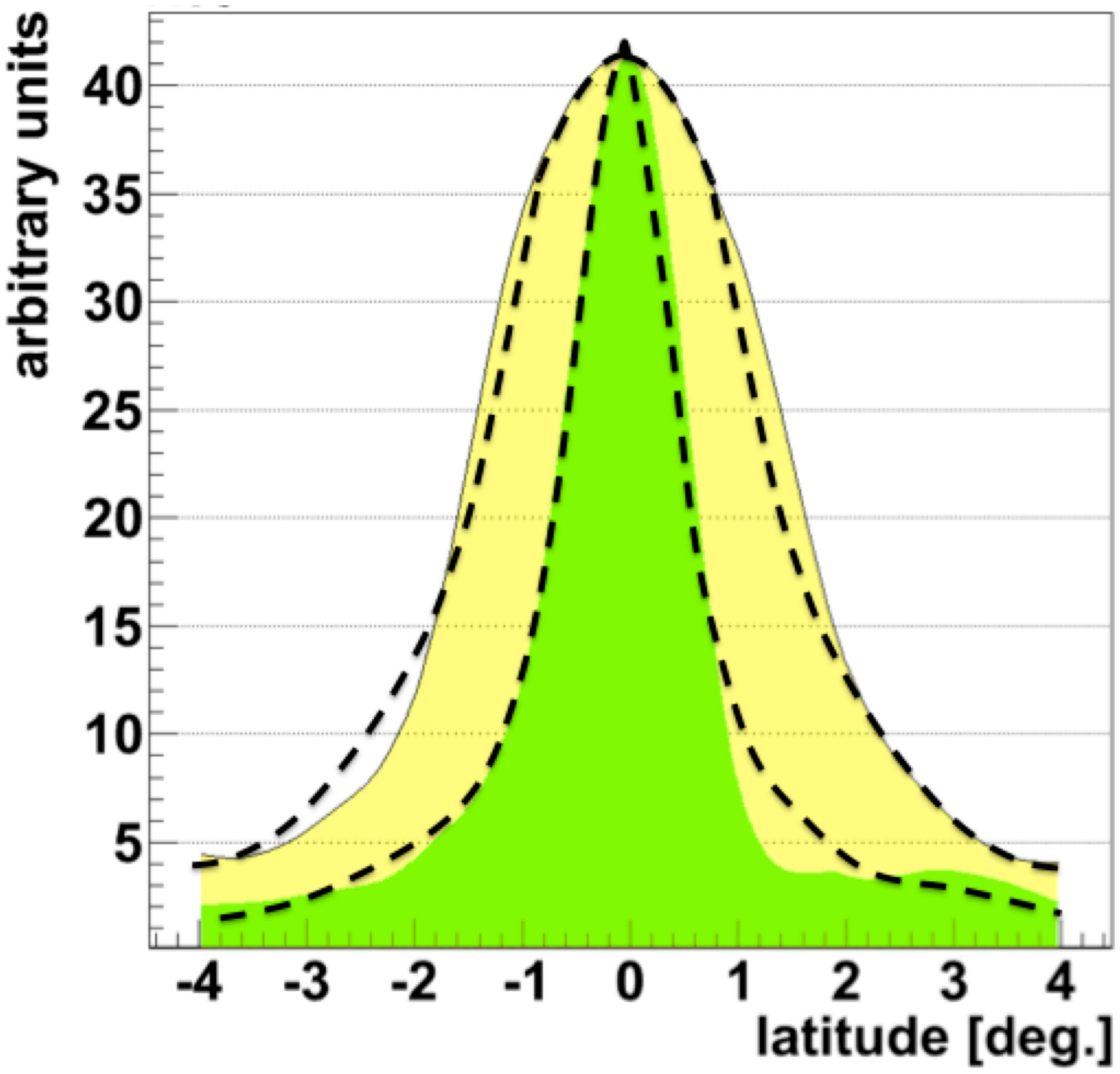}\hspace*{3mm}
\includegraphics[width=0.385\textwidth,height=0.5\textwidth,clip]{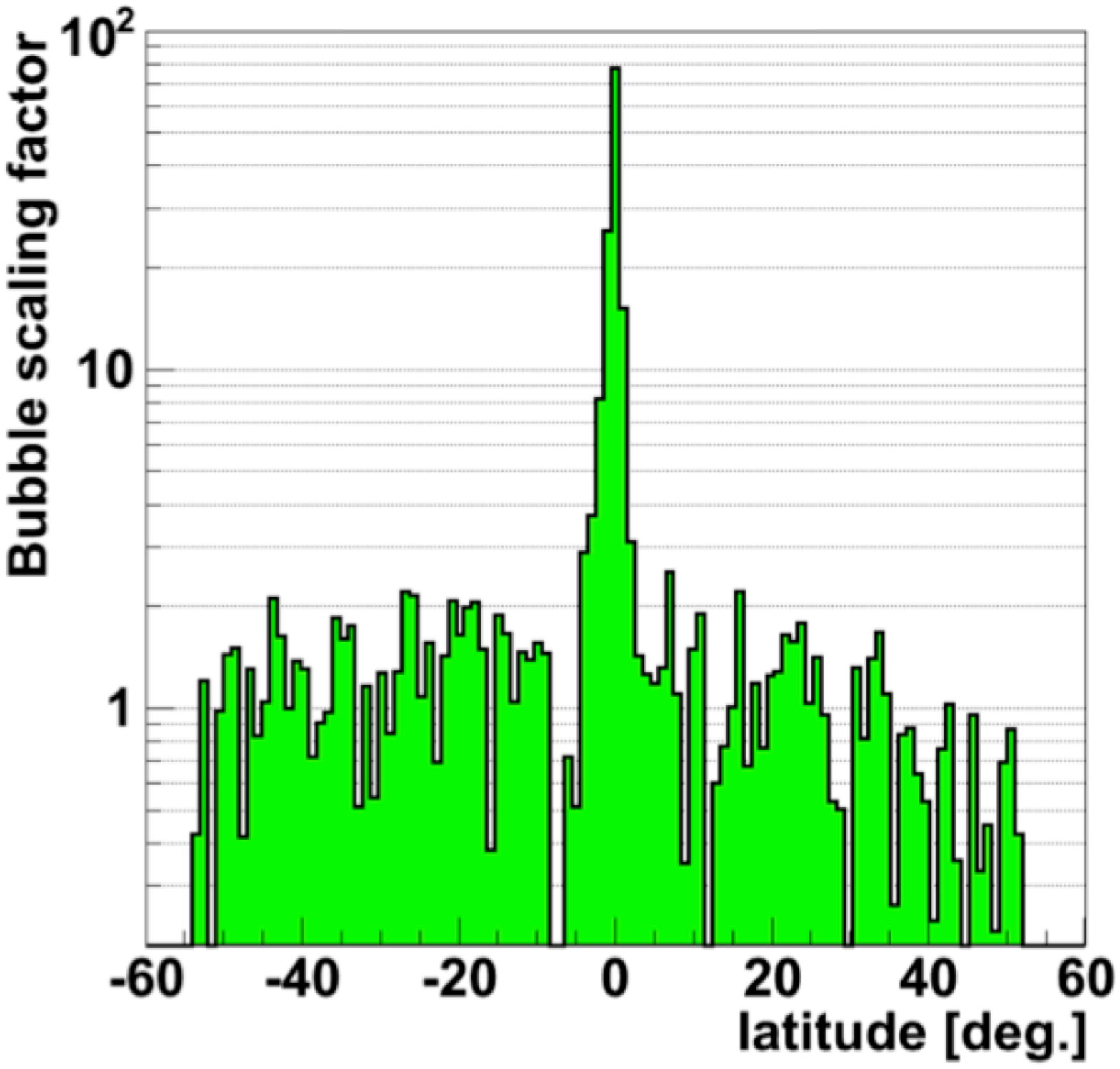}\\[-16mm]
\hspace*{0.03\textwidth}(e)\hspace*{0.35\textwidth} (f)\\[-1.2mm]
\caption{(a) The results from a template fit without Bubble template to the gamma-ray spectrum in the Galactic plane for longitudes centred around 35$^\circ$.
(b) After adding the Bubble template to the fit, the reduced $\chi^2/n_{dof}$ decreases from 44.1/19=2.3  (left) to 3.6/18=0.2.  (c) and (d) The template fits to the southern Bubble and inner Galactic Plane including the point sources.  (e) Overlaid latitude distributions for the  diffuse (outer, yellow) and bubble-like emission (inner, green)   for $|l|<20^\circ$. 
The distributions are normalised at the centre and the  dashed lines show the expected distributions. 
(f)  Latitude distribution of the Bubble for $|l|<20^\circ$ with  contributions up to $|b| < 55^\circ$ and a large enhancement in the Plane.
 \label{f2}
}
\end{figure}

The foreground templates  depend mainly on the CR spectral shapes. We tuned the CR spectra of electrons and protons and other nuclei to best describe the locally observed CR spectra and the gamma-ray sky. 
For this purpose we used the public propagation code Dragon \citep{Evoli:2008dv}, which  uses up-to-date accelerator data for the cross sections, considers variations of the magnetic and interstellar radiation field and takes CR energy losses and kinematic aspects for the production of gamma-rays into account. Therefore it is a convenient tool to determine the spectral shapes in different sky directions. 
The resulting foreground templates are shown in Fig. \ref{f1}a-\ref{f1}c. Here we superimposed the templates in the 1025 different directions discussed above. All spectra were normalised at 3.4 GeV, so only differences in shape, not in flux, are shown. One observes that the $\pi^0$ template does not change with direction, as expected, since the diffusion time of protons or other nuclei is faster than the energy loss time in the energy range of interest. 
For bremsstrahlung  the negligible differences are due to the varying magnetic field in the Galaxy. For inverse compton  the differences can be up to 30\% because of the difference in the spectra of the ISRF, which is dominated by the CMB background in the halo, but has contributions from infrared  and stellar light in the disc.  These differences in different directions were taken into account in the template fit. Similar templates are obtained from the Galprop public propagation code \citep{Moskalenko:1998id,Vladimirov:2010aq}, but the the well-tuned Dragon code provides a considerably better fit.
\begin{figure}
\centering
\includegraphics[width=0.45\textwidth,clip]{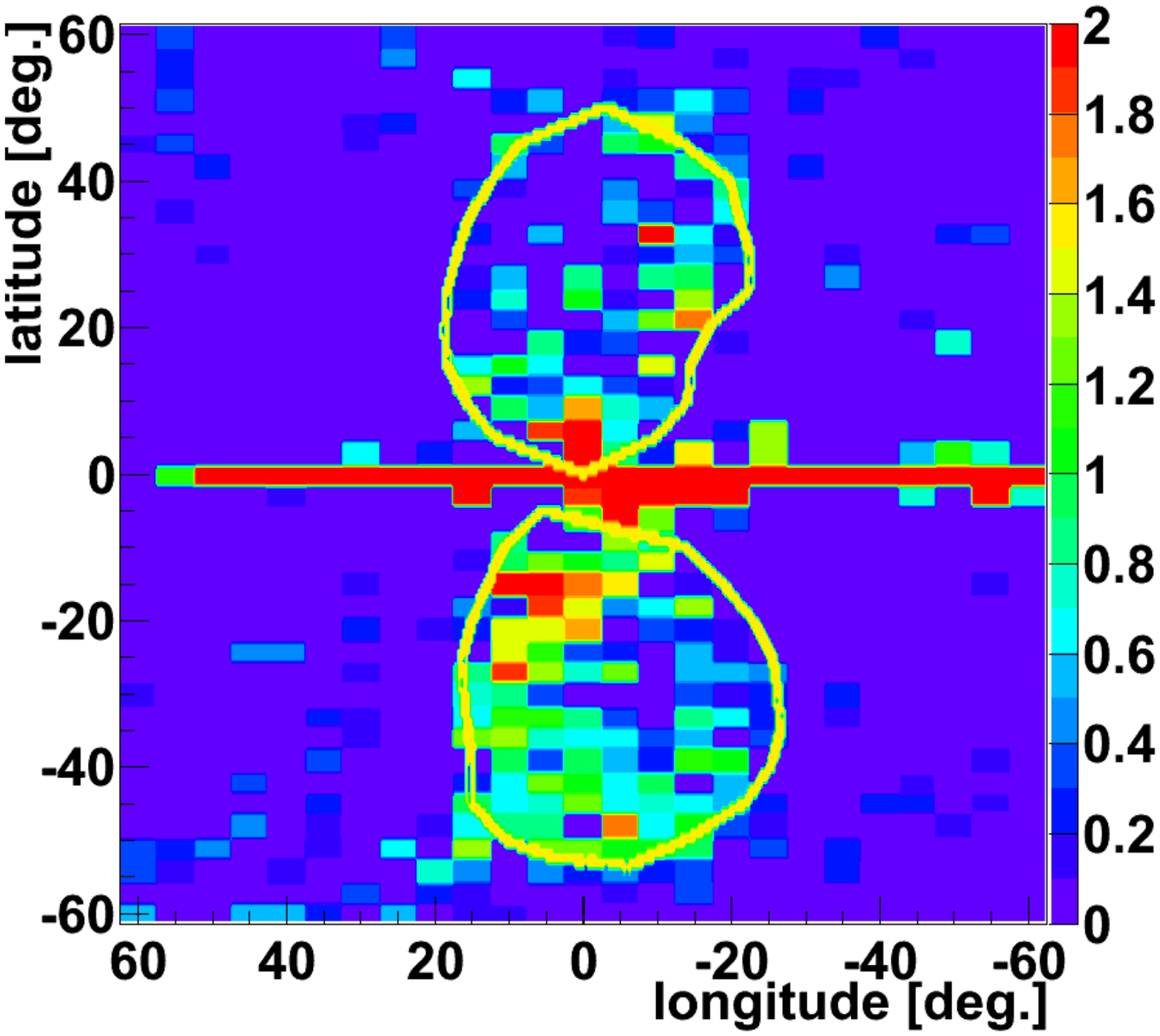}
\includegraphics[width=0.45\textwidth,clip]{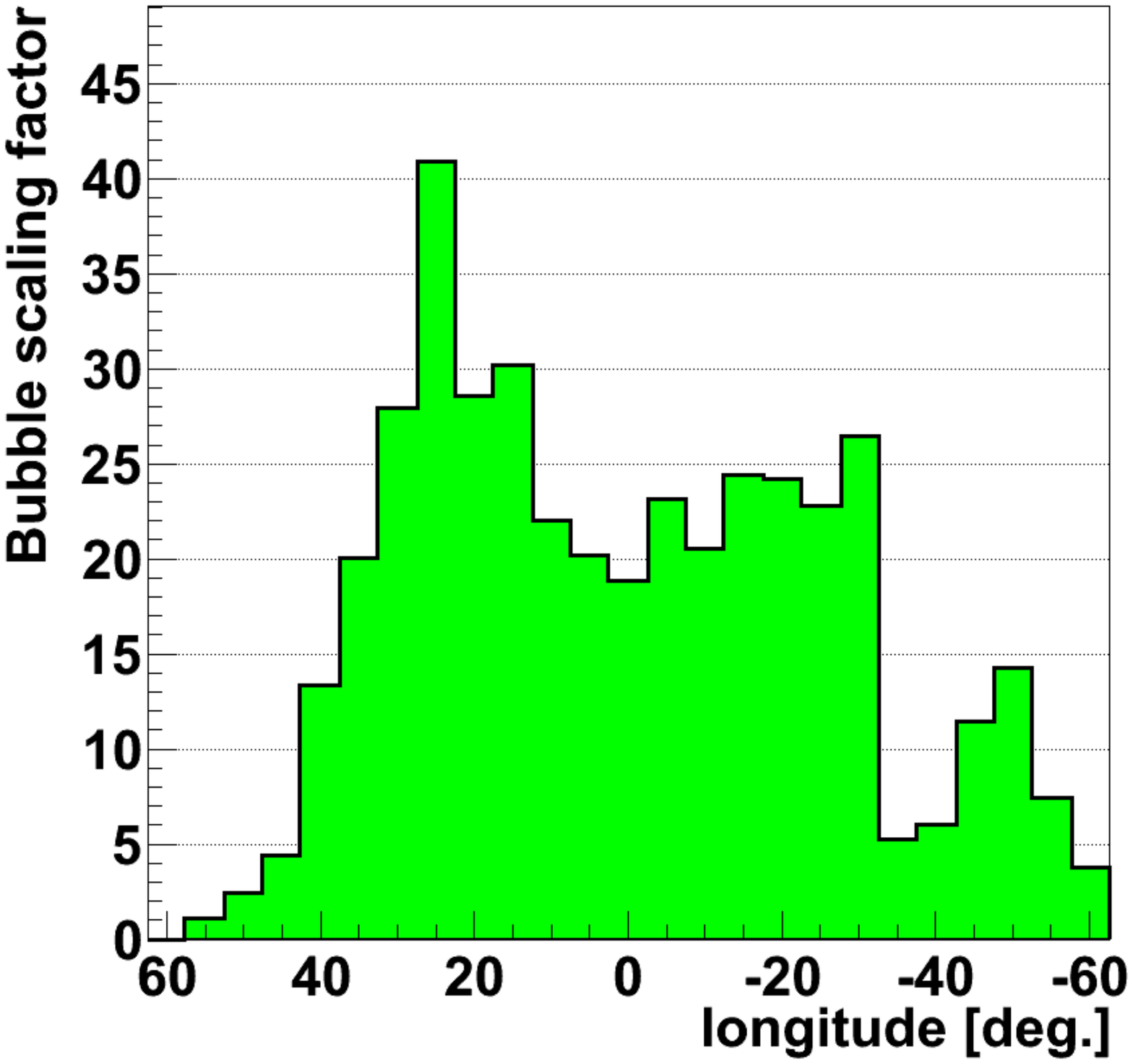}\\[-15mm]
\hspace*{0.01\textwidth}(a)\hspace*{0.4\textwidth} (b)\\[5mm]
\caption{(a) Intensity of the Bubble template in the $l,b$ plane.  The surrounding line  corresponds to the morphology from \citet{Su:2013}, who find a homogeneous flux inside the line.  The units correspond to the Bubble template unit in Fig. \ref{f1}d. (b)  Longitude distribution for the bubble-like intensity in the Galactic Plane ($|b|<3^\circ$). 
 \label{f3}
}
\end{figure}
Towards the GC the isotropic contribution is small and since its normalisation is fixed mainly by regions outside our region-of-interest and the normalisation is the same for all subcones, its normalisation is fixed in our fit. Varying the isotropic component by $\pm 10\%$  from its nominal FST value worsened the fit.

We  take the normalisations of the BR and IC templates  to be 100\% correlated, since they are both proportional to the electron number density. The ratio $n_3/n_2$  of the two templates is given by the ratio of  cross sections times target densities (gas and ISRF), which was taken from the Dragon program for each sky direction. So in total we  have only 2 free parameters to describe the foreground in each subcone, namely  $n_1$ and $n_2$. 

An example of a template fit with foregrounds only is shown in Fig. \ref{f2}a for a single subcone. One observes that the data below 20 GeV are well described by the sum of the foregrounds. However, above 20 GeV the data show an excess. This is the observation of the bubble-like emission. Including the Bubble template leads to the fit shown in Fig. \ref{f2}b. The indicated Bubble template  corresponds to the CR spectra expected  from SCRs with a spectral index of $2.15\pm 0.1$, as expected from diffusive shock wave acceleration \citep{Hillas:2005cs}- The depletion below 1 GeV  originates from the kinematics of $\pi^0$ production.  This spectrum  does not fall as sharply  at low energies as found previously \citep{Su:2010qj,Su:2013}, see Fig. \ref{f1}d.  Such a fast depletion would  require a sharp break in the underlying proton spectrum at  20 GV, but such a break worsens the fit. Therefore, we use a CR power law with a spectral index of 2.15 without any break, both for nuclei and electrons. The resulting IC and BR spectra are shown in Fig. \ref{f1}d as well. They did not result in good fits. 
%
\begin{figure}
\centering
\includegraphics[width=0.6\textwidth,clip]{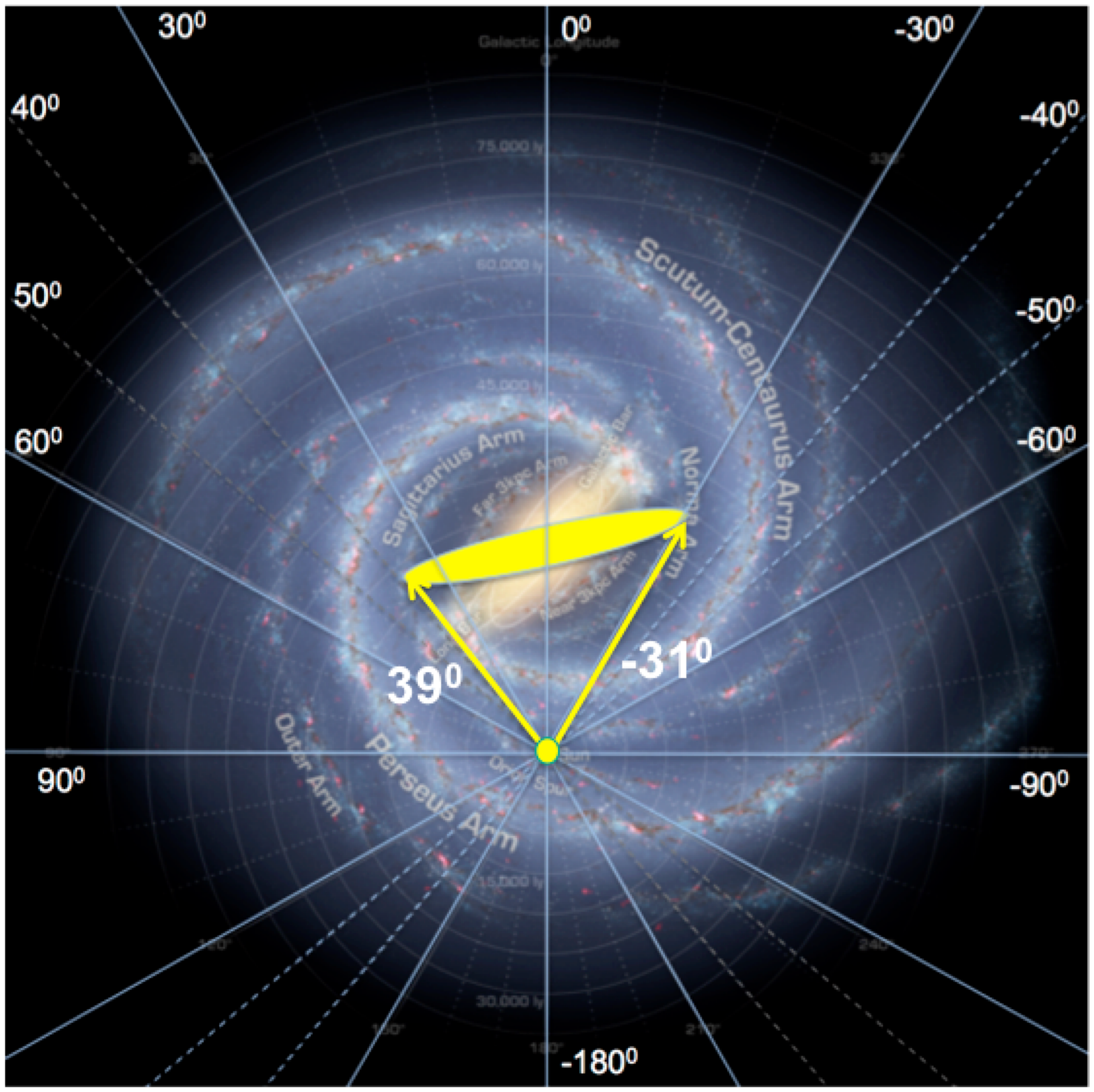}
\caption{Sketch of the Milky Way with the central Bar and spiral arms. The picture was adapted from  \citet{Churchwell:2009zz}). The "slim (yellow) ellipse" in the centre indicates the angle and length of the Bar obtained from the endpoints of the bubble-like emission in Fig. \ref{f3}b.
 \label{f4}
}
\end{figure}

The  template fit discussed above was repeated for all 1025 subcones. Surprisingly,  with only 3 free parameters for each direction ($n_1, n_2, n_4$) we can describe the gamma-ray sky in any direction with an excellent    $\chi^2$.  A few regions with a high $\chi^2$ (i.e. a poor fit) at high latitudes  have been excluded. Inspection of the spectra showed that the data are below  the isotropic flux, which is attributed to imperfect subtraction of a clustering of point sources in these subcones. Removing these regions reduced our region-of--interest by 8\%.  Figs. \ref{f2}c and \ref{f2}d show the tempate fits for the south Bubble and the central Galactic Plane  without point source subtraction. The north Bubble spectra look similar. The normalisation of the point sources was taken from the second Fermi catalogue. One observes a tail in the point sources similar to the shape of the Bubble template, as expected if the bubble-like contribution is related to unresolved SNRs. These are expected predominantly in molecular clouds. Since the latter have a narrower latitude distribution than the HI  gas component, one expects for the bubble-like emission a narrower latitude distribution in comparison with the diffuse emission.
This is indeed the case as shown  in Fig. \ref{f2}e, where we compare the latitude distributions of the $n_1$ and $n_4$ components. The latter component has a FWHM of 175 pc and follows closely the latitude distribution from the molecular clouds, as shown by the inner dashed line taken from \citet{Dame:2000sp}.  The  HI layer has a wider distribution of 230 pc FWHM \citet{Dickey:1990mf}.  To obtain the width in latitude of the diffuse gamma-rays one has to weigh the emissivity of the sum of the gas layers with the CR density. The corresponding weight factor along the line-of-sight for a given latitude bin was taken from the fit assuming the gas distributions mentioned above. The resulting distribution is shown by the outer dashed line.

The intensity ratio R of the SRCs and diffuse Galactic CRs from $\pi^0$ production was estimated by \citet{2004ApJ...611...12B}:
 \begin{equation} R(E)=0.07\left(\frac{N_g^{SCR}}{N_g^{GCR}}\right)\left(\frac{T_p}{10^5 \ yr}\right)\left(\frac{E}{1\ GeV}\right)^{0.6}, \label{e3}\end{equation} where the first bracket takes into account the difference in gas densities in SNRs and the Galactic disc, the second bracket the limited confinement time $T_p$ of SCRs in SNRs and the third bracket the difference in energy dependence between the Bubble template and the $\pi^0$ template.
From Figs. \ref{f2}b and \ref{f2}d one observes that this ratio becomes one for energies between 15 and 30 GeV, which requires that the product of the first two brackets is of the order of 4. This is a reasonable number given that the gas density in molecular clouds is higher than in the disc.

\section{Morphology of the Fermi Bubbles}
The fitted values of  $n_4$ in all subcones are plotted in Fig. \ref{f3}a, which obviously show a similar morphology as found  by  \citet{Su:2010qj}  (indicated by the surrounding line), but we find a much richer structure inside because of the  better spatial resolution with our method..  However, the surprise is a  strong bubble-like  emission in the  disc,  shown by the intense (red)  bar in the centre of  Fig. \ref{f3}a. Its distribution as function of longitude is shown in Fig. \ref{f3}b, which reveals  a strong increase at $l=+39^\circ$  followed by a sharp decrease at $l=-31^\circ$. These angles are close  to the endpoints of the Bar, as pictured in Fig. \ref{f4}. Assuming these to be the endpoints  and taking a distance between the GC and the Sun of $8.3\pm 0.4$ kpc \citep{Genzel:2010zy} the morphology of the Bar is completely determined, as is apparent from the geometry shown in Fig. \ref{f4}. Using a conservative error for the directions to the endpoints of $\pm 1^\circ$ we find from a fit that the major Bar axis makes an angle of  $77.7^\circ \pm 2.1^\circ$  with respect to the GC-Sun line and has a half length  of 5.9$\pm$0.1 kpc. This morphology is shown by the slim (yellow) ellipse in Fig. \ref{f4}. It has a  larger Bar angle than the Bar indicated in the background, which has  $44^\circ\pm 10^\circ$ \citep{Churchwell:2009zz}, but this angle has a large error, since it was found from a subset of stars in the first quadrant.

We find additional evidence for the interpretation of the bubble-like component as being due to SCRs  from the fact that we see the bubble-like emission also at longitudes around  -50$^\circ$ as a separate peak in Fig. \ref{f3}b.
\begin{figure}[t]
\centering
\includegraphics[width=0.32\textwidth,clip]{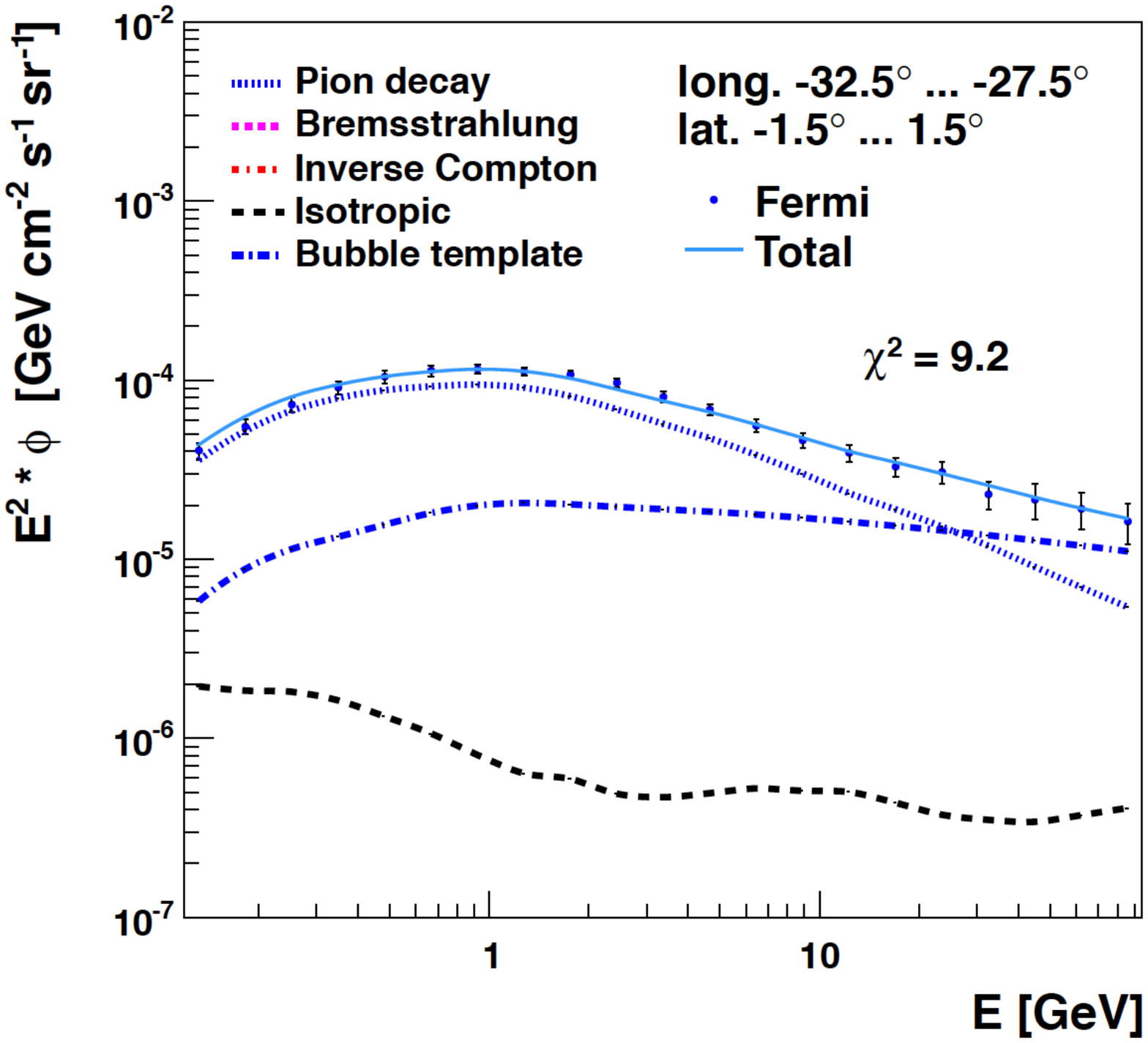}
\includegraphics[width=0.32\textwidth,clip]{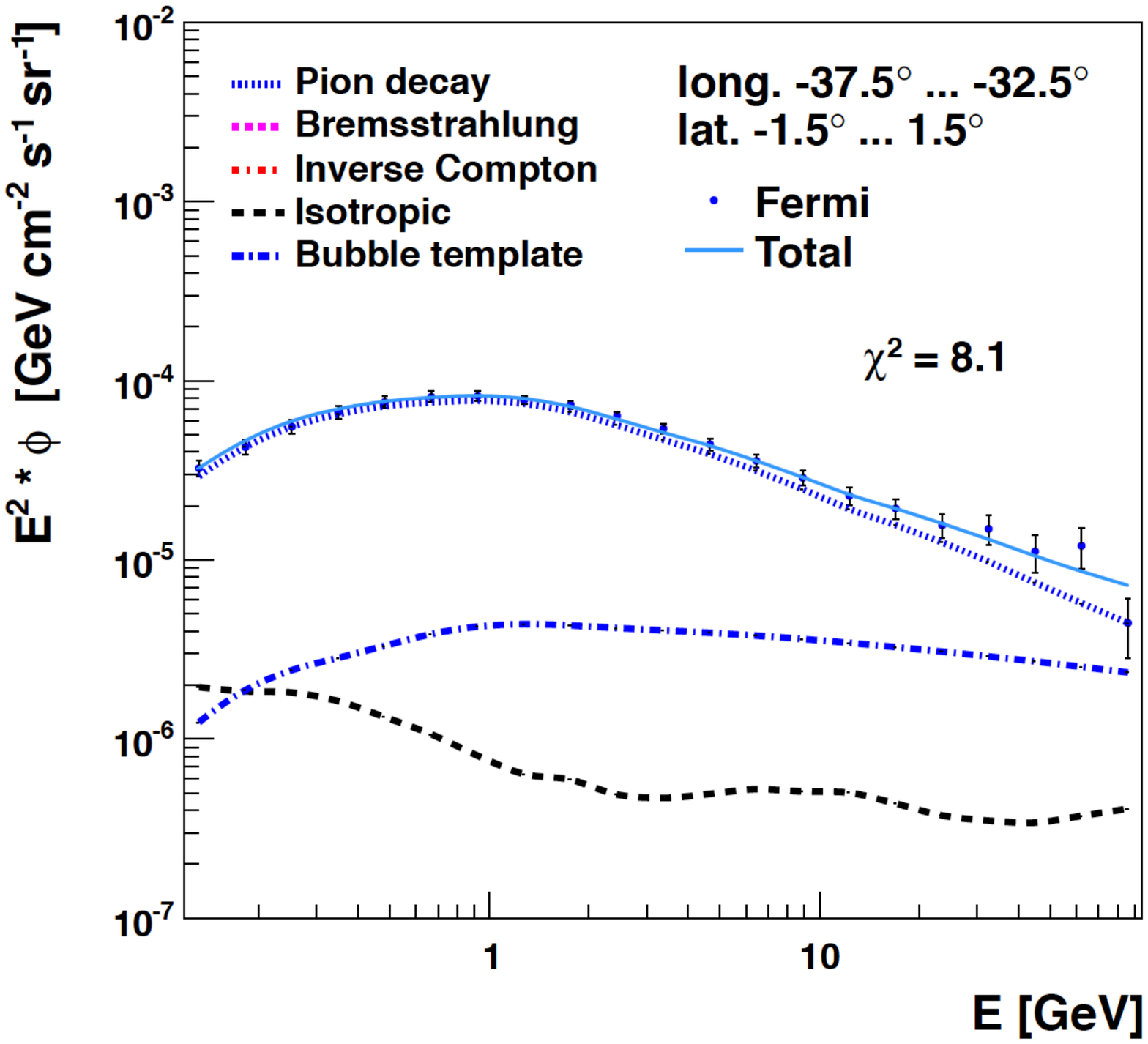}
\includegraphics[width=0.32\textwidth,clip]{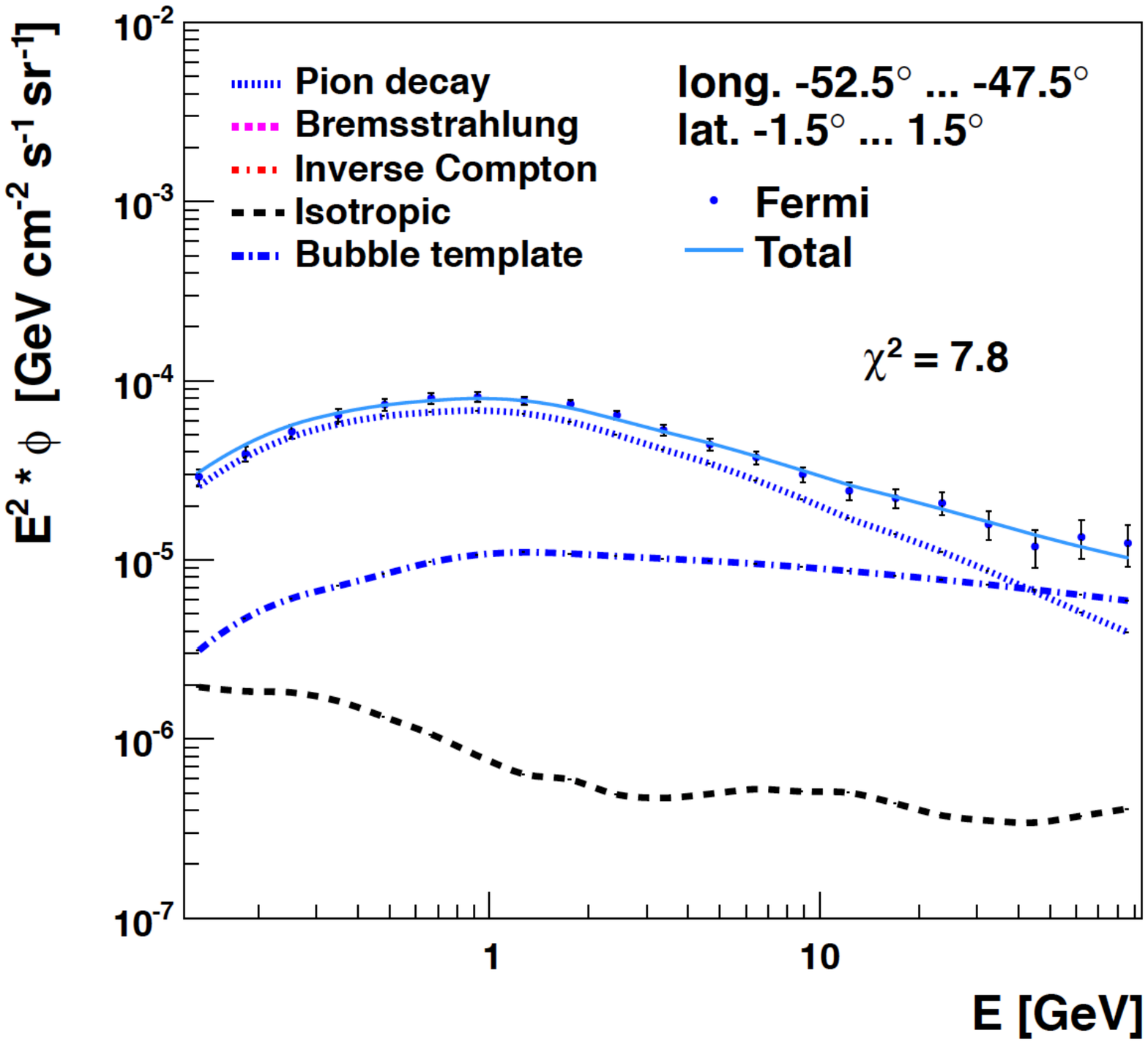}\\[-15mm]
\hspace*{0.02\textwidth}(a)\hspace*{0.30\textwidth} (b)  \hspace*{0.30\textwidth} (c) 
\caption{From left to right: The results from a template fit to the gamma-ray spectrum in the Plane at longitudes centred at   -30$^\circ$, -40$^\circ$, -50$^\circ$, respectively. One observes clearly a decrease of the bubble-like emission near $l=-40^\circ$, which corresponds to the dip in Fig. \ref{f3}b. 
\label{f5}
}
\end{figure}
 This is exactly the region of the tangent point of the Scutum-Centaurus spiral arm, as shown in Fig. \ref{f4}.  The evidence for this dip is demonstrated in Fig. \ref{f5}:  one observes  a strong bubble-like emission in the far end of the  Bar ($l=-30^\circ$), a lower contribution at the tangent point of the Scutum-Centaurus arm  ($l=-50^\circ$) and a still smaller contribution in between ($l=-40^\circ$), where the arm is further away.

We interpret the combined findings as follows:  supernovae explosions accelerate CRs  with a spectral index of 2.15 by diffusive shock wave acceleration, which in turn produce gamma-rays with such a hard spectrum as long as they are trapped in the sources (SCRs).   In the central region ($|l|<20^\circ$) the global thermal and CR pressure are assumed to be high enough to blow a small fraction of the gas into the halo, either from the disc \citep{Everett:2007dw,Breitschwerdt:2008na} or from the inner GC \citep{Crocker:2011en,Crocker:2010qn}. This hot gas in the halo is inferred from the ROSAT X-ray data  \citep{Snowden:1997ze}, which was interpreted as advection of gas from the Plane with wind speeds as large as 2000 km/h \citep{BlandHawthorn:2002ij}.

In such an advective environment the CRs can be trapped inside the plasma and there is no energy-dependent escape, so the  SCRs will still correspond to the 1/$E^{2.15}$  spectrum at high latitudes.  The spectral index of the locally observed proton spectrum is around 2.75 and the softening from 2.15 to 2.75 is attributed to diffusion, simply because high energy protons escape faster from  the Galaxy \citep{Strong:2007nh}. 

 The luminosity of the Bubbles   between 1 and 100 GeV for $10^\circ<|b|<55^\circ$ and  $|l|<30^\circ$ corresponds to $5.6\pm0.3(fit)\pm0.9(sys)\cdot10^{37}$   erg/s. Here we followed  the calculation  in the discovery paper by \citet{Su:2010qj}, who found  $4\cdot10^{37}$ erg/s   without giving an error. The first error originates from the fit, while the second error originates from the error in the spectral index, which is the dominant systematic uncertainty because the normalisation of the Bubble template is most sensitive to the high energy tail of the data, so the extrapolation to lower energies depends on the spectral index. The Bubble luminosity increases to  $8.7\pm0.4(fit)\pm1.5(sys)\cdot10^{37}$ erg/s, if we decrease the  latitude limit to  $1.5^\circ$, while the bubble-like emission in the Plane ($|b|<1.5^\circ$) has a luminosity of $6.9\pm0.6(fit)\pm1.1(sys)\cdot10^{37}$ erg/s. This can be compared with the hadronic energy release from SNRs in the inner Galaxy.  We expect for the luminosity in gamma-rays  between 1 and 100 GeV:  $E_\gamma^{SCR}·=\epsilon_{CR}\  \epsilon_\gamma\  \epsilon_{(1-100)} \ \epsilon_{SCR/(SCR+GCR)}\ E_0$ erg/s. Here  $\epsilon_{CR}$ is the fraction of the SNR mechanical power $E_0$ into hadronic CR energy,  $ \epsilon_\gamma$ is the energy converted into gamma-rays, $ \epsilon_{(1-100)}$ is  gamma-ray energy fraction between 1 and 100 GeV, and $\epsilon_{SCR/(SCR+GCR)}$ is the fraction of energy into SCRs.
\citet{Hillas:2005cs} estimates 1 SNR/century in the inner region ($R<4$ kpc) corresponding to $E_0=10^{51}\ erg/3.10^9\ s=3.3\cdot 10^{41}$ erg/s. Using   $\epsilon_{CR}\ \epsilon_\gamma\  \epsilon_{1-100}\ \epsilon_{SCR/(SCR+GCR)}$=0.16 x 0.1 x 0.3 x 0.3=0.0009 we find  $E_\gamma=4.8\cdot10^{38}$ erg/s  for the gamma-ray energy from SCRs to be compared with a summed bubble-like emission in the halo and disc of $E_\gamma^{SCR}=1.6\cdot10^{38}$ erg/s. This is reasonable agreement given the large uncertainties involved.
We obtained $\epsilon_{SCR/(SCR+GCR)}=0.3$ from the energies in the $n_4$ and $n_1$ components in the fit, while the other efficiencies were estimated as in \citet{Hillas:2005cs}.

In summary, with our template fitting method we confirm the Bubbles in the halo, but find in addition  strong bubble-like emission in the star-forming regions of the Galactic Plane, thus excluding many  models  mentioned in the introduction,  like dark matter annihilation or jet-activity from the Galactic centre as the sole source of the hard gamma-ray component. The latitude distribution of the bubble-like emission shows that it is  strongly correlated with the molecular gas distribution, as expected for SCRs and proving its hadronic origin The hadronic origin is also apparent from the observed spectral shape, as shown in Fig. \ref{f1}d. The morphology of the bubble-like emission with star-forming regions combined with an underlying proton spectrum proportional to 1/$E^{2.15\pm0.1}$  provides strong evidence, that these are the predicted contributions from SCRs \citep{2004ApJ...611...12B}. The sharp edges of the bubble-like emission in the Plane coincide with the endpoints of the Galactic Bar. Taking these edges as endpoints of the Bar we  find the morphology of the Bar to be consistent with previous measurements, but with a higher precision (Bar angle 77.7$ \pm 2.1^\circ$,  half length   5.9$\pm$0.1 kpc), since gamma-rays above a few GeV are hardly absorbed by dust. 
 \citet{Kretschmer:2013naa} find an outflow of $^{26}$Al, which is a radioactive element from SNRs, in exactly  the longitude range of our Bar parameters, although they presume it to be from spiral arms. 
However, their linearly-increasing rotation speeds up to  $\pm$300 km/s coincide with  our Bar length of 5.9 kpc and a Bar pattern speed  of $53\pm3$  km/s/kpc \citep{Dehnen:1999fi}.

\acknowledgments
Support from the Deutsches Zentrum f\"ur Luft- und Raumfahrt (DLR)   and helpful discussions with Roland Crocker and Heinz V\"olk are warmly  acknowledged.
 We thank Simon Kunz for providing  optimized  Dragon parameters and  the referee for detailed questions, which greatly improved the manuscript. We are grateful to the Fermi scientists, engineers and technicians for collecting the Fermi data and the Fermi Science Support Center for providing strong support for innocent guest investigators.


\end{document}